\documentclass[
]{ceurart}

\begin{document}

\copyrightyear{}
\copyrightclause{}

\conference{ }

\title{AI Security Threats against Pervasive Robotic Systems: \\ A Course for Next Generation Cybersecurity Workforce}

\author[]{Sudip Mittal}[%
]

\author[]{Jingdao Chen}[%
]

\address[]{Department of Computer Science \& Engineering, \\
  Mississippi State University, Mississippi State, MS, USA \\ 
  email: mittal@cse.msstate.edu, chenjingdao@cse.msstate.edu}

\begin{abstract}
  Robotics, automation, and related Artificial Intelligence (AI) systems have become pervasive bringing in concerns related to security, safety, accuracy, and trust. With growing dependency on physical robots that work in close proximity to humans, the security of these systems is becoming increasingly important to prevent cyber-attacks that could lead to privacy invasion, critical operations sabotage, and bodily harm. The current shortfall of professionals who can defend such systems demands development and integration of such a curriculum. 

  This course description includes details about seven self-contained and adaptive modules on ``AI security threats against pervasive robotic systems''. Topics include: 1) Introduction, examples of attacks, and motivation; 2) - Robotic AI attack surfaces and penetration testing; 3) - Attack patterns and security strategies for input sensors; 4) - Training attacks and associated security strategies; 5) - Inference attacks and associated security strategies; 6) - Actuator attacks and associated security strategies; and 7) - Ethics of AI, robotics, and cybersecurity. 
\end{abstract}

\begin{keywords}
  Cybersecurity \sep
  Robotics \sep
  Artificial Intelligence \sep
  Adversarial Artificial Intelligence
\end{keywords}

\maketitle

\section{Introduction}

{Robotics, automation, and related Artificial Intelligence (AI) systems} have become {pervasive} in our daily lives and transformed operations. Modern households, construction sites, warehouses, hospitals, precision agriculture, military, emergency workers, and more, use different sets of robots to provide workflow augmentation and mobility assistance. Robotic technologies have experienced {mass adoption in the consumer space, industry, and critical infrastructures} bringing in concerns related to security, safety, accuracy and trust \cite{morante15}. 
The International Federation of Robotics estimates that in 2021, there is total of{ 3 million industrial robots operating in factories} around the world \cite{ifr}, manufacturing vehicles, assembling electronics, and packaging food that we use in our daily lives.
On a similar note, McKinsey estimates that autonomous vehicles (AVs) in urban areas could constitute a {\$1.6 trillion market in 2030} \cite{mckinsey}.
However, {mass adoption of robotics technology} in applications such as autonomous vehicles, drones, and domestic robots can also {leave the general public vulnerable to physical harm and damage when such robotic systems are compromised}.
In 2015, a Jeep Cherokee was remotely hacked while being driven, causing Fiat Chrysler to recall 1.4 million vehicles and pay \$105 million in fines to the National Highway Traffic and Safety Administration \cite{jeepcherokee}.
In 2016, a {malfunctioning security robot} at a mall in California knocked a 16-month-old child to the ground and ran over one of his feet \cite{robottoddler}.
Robotic systems tend to make use of large, interconnected stacks of software handling sensing, communication, planning, and actuation, any of which can be an attack surface for a malicious agent \cite{alurity2020}. In addition, robots that use the commonly-adopted ROS framework are also subject to vulnerable communication between software nodes \cite{teixeira2020}.



With growing dependency on physical robots that work in close proximity to humans, the {security of these systems} is becoming increasingly important to prevent cyber-attacks that could lead to invasion of personal privacy, sabotage of critical operations, and even bodily harm. 
Two high profile research papers by Eykholt et al. and  Sharif et al. highlight the security issues with AI models deployed in real world robots and their subsystems \cite{eykholt2018robust,sharif2016accessorize}. Sharif et al. demonstrated that a person wearing crafted glasses can trick facial recognition systems. Eykholt et al. applied adversarial stickers to road signs causing them to be misinterpreted by an autonomous vehicle (a robot). These examples indicate that various learning systems, especially robots, remain vulnerable to a new class of attacks, when an adversary attempts to tamper with them or interact with them maliciously.

In this course, targeted at senior undergraduate and graduate students, we aim to {create a cohort of technically trained students that have skills in pervasive intelligent robotic security}. 
Providing students with this education will be an incredibly powerful tool to bridge the cybersecurity talent gap. It will open up the opportunities for not only cybersecurity focused talent, but also from students across other concentrations like data science or AI. On the other side, cybersecurity focused students who develop an understanding of these concepts can expect to open many more opportunities in this highly sought-after field.

To {gauge the level of understanding about security threats to robots and their AI subsystems} among {\textit{incoming} university first-year and fourth-year students}, we conducted {an initial survey} study in August 2021 and August 2022. The students were informed about the attack presented by Eykholt et al. \cite{eykholt2018robust}. {86\% of the incoming freshman class were ``surprised''} by the possibility of such an attack, and responded that such attacks against an autonomous vehicle can potentially impact its usability. {90\% of the fourth-year computer science students}, who responded {do not believe} that they possess the {required skills to check for exploitable AI vulnerabilities} in models they create; and {96\% of them do not know any security strategies} that can protect a robot's (or an autonomous vehicle as described by Eykholt et al. \cite{eykholt2018robust}) AI subsystems. {These results} further emphasize the need for education and awareness in students about this critical domain. {\textit{Inculcating a culture of preparedness in the future workforce about AI security threats to pervasive robotic systems is important to prevent serious social and economic implications}}. 

In this paper, we describe a course on ``AI Security Threats to Pervasive Robotic Systems'' to transfer cutting edge research in this domain to senior undergraduate and graduate students. This course will raise awareness among the target population, to the fact that robotic AI systems are systematically vulnerable to attacks. After which it will provide them with defensive strategies to prevent and protect these systems from attacks. The course {fulfills an educational niche} by specifically addressing the evolving space of cybersecurity threats involved with modern AI and robotic systems, a topic not covered by most existing courses in robotics and cybersecurity.

This paper is organized as follows: Section \ref{over} provides a course overview and the required pre-requisites. The detailed description of the differnt course modules is available in Section \ref{ucurr}. We present our conclusion in Section \ref{conc}. 

\section{Course Overview \& Pre-requisites}\label{over}

\begin{figure}[h]
    \centering
    \includegraphics[scale=0.43]{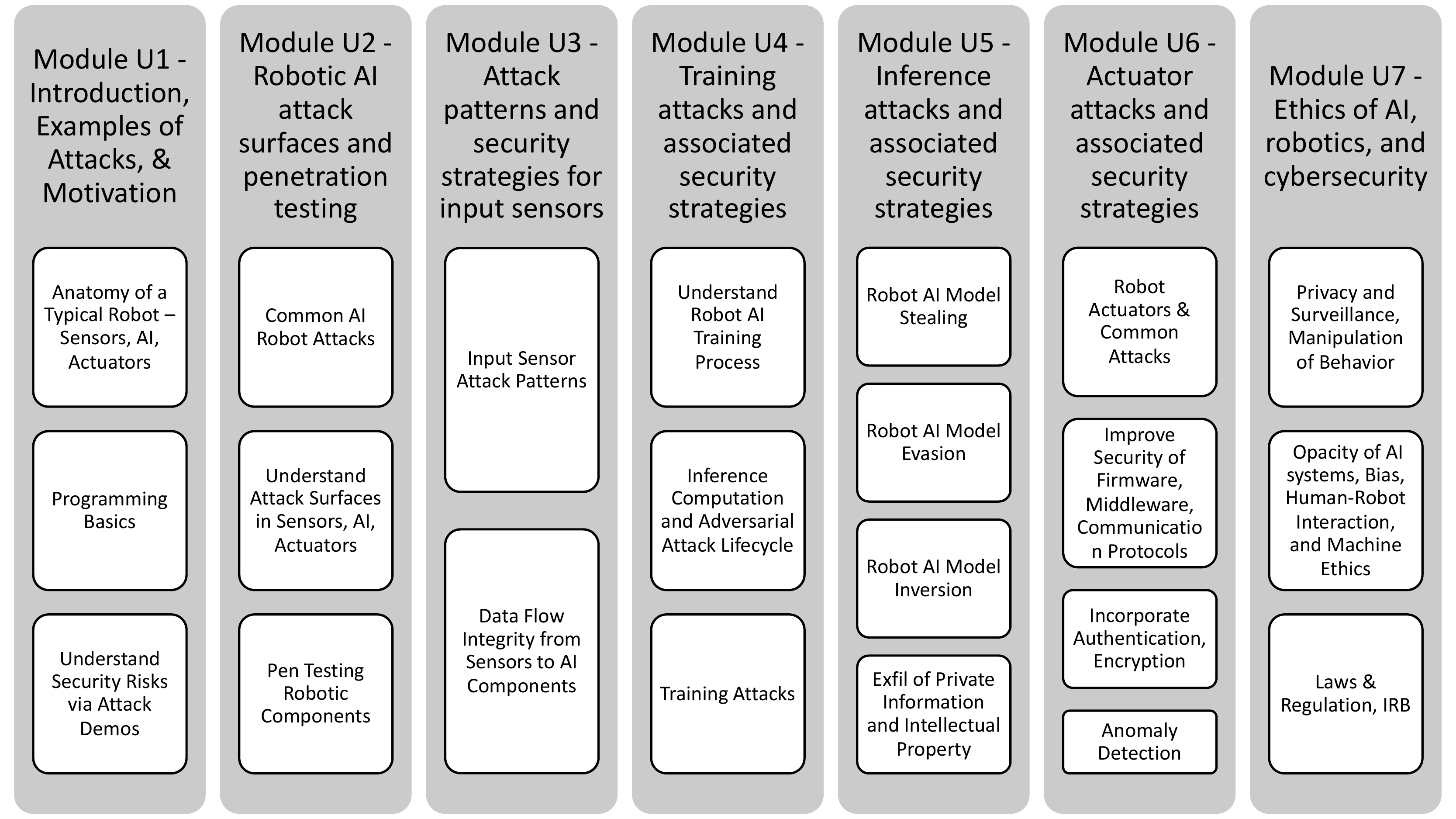}
    \caption{Modules and Major Learning Outcomes.}
    \label{fig:overview}
\end{figure}

Our {curriculum} (See Section \ref{ucurr}), will have {7 modules}, including, U1 - Introduction, Examples of Attacks, \& Motivation; U2 - Robotic AI attack surfaces and penetration testing; U3 - Attack patterns and security strategies for input sensors; U4 - Training attacks and associated security strategies; U5 - Inference attacks and associated security strategies; U6 - Actuator attacks and associated security strategies; and U7 - Ethics of AI, robotics, and cybersecurity. An overview of the course along with major learning outcomes is available in Figure \ref{fig:overview}.

Before enrolling in a course that teaches the proposed modules, students should have completed the listed {pre-requisites}:

\begin{enumerate}
    \item \textit{Introduction to Machine Learning/Artificial Intelligence} ({required}) - Students are supposed to have taken an undergraduate machine learning/artificial intelligence course, generally taught to computer science majors at various universities. 
    \item \textit{Introduction to Robotics} ({recommended but not necessary}) - This class should introduce students to the basic components of robots including sensors, actuators, and relevant software libraries. Experience with robotic capabilities like, sensing, planning, control, and learning, operating under real-world conditions will also be beneficial. 
    \item \textit{Introduction to Cybersecurity} ({recommended but not necessary}) - The proposed course expect students to have basic cybersecurity foundations, which are completed in 3 foundational knowledge units (Cybersecurity Foundations, Cybersecurity Principles, IT Systems Components) as discussed in the  NSA/DHS CAE-CDE designation requirements \cite{caecde}.
\end{enumerate}

\section{Course Module Descriptions}\label{ucurr}

Here we present a summary of topics part of our proposed course. This course will have {7 modules}, U1 - Introduction, Examples of Attacks, \& Motivation; U2 - Robotic AI attack surfaces and penetration testing; U3 - Attack patterns and security strategies for input sensors; U4 - Training attacks and associated security strategies; U5 - Inference attacks and associated security strategies; U6 - Actuator attacks and associated security strategies; and U7 - Ethics of AI, robotics, and cybersecurity. {Each module consists of lectures and lab sessions}.

To make the course hands-on and project centric we suggest a demonstration kit along with some open source emulators that can be used to build lab assignments. 

\begin{figure}[h]

  \begin{center}
        
    \includegraphics[width=0.33\textwidth]{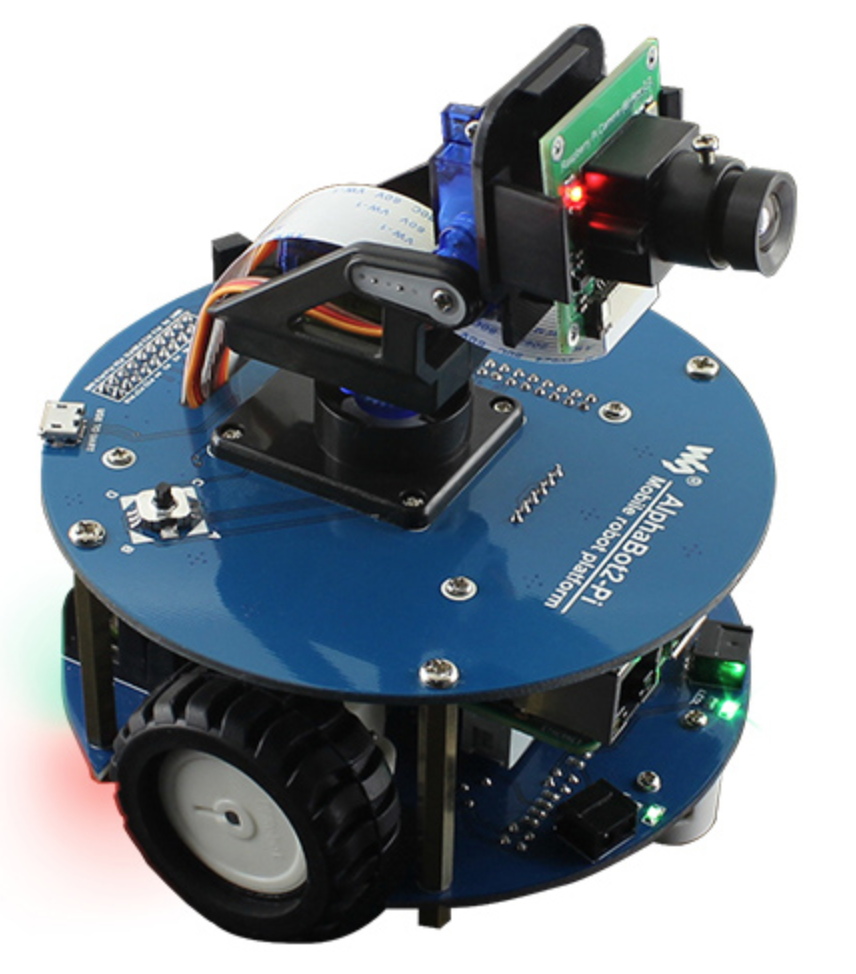}\includegraphics[width=0.45\textwidth]{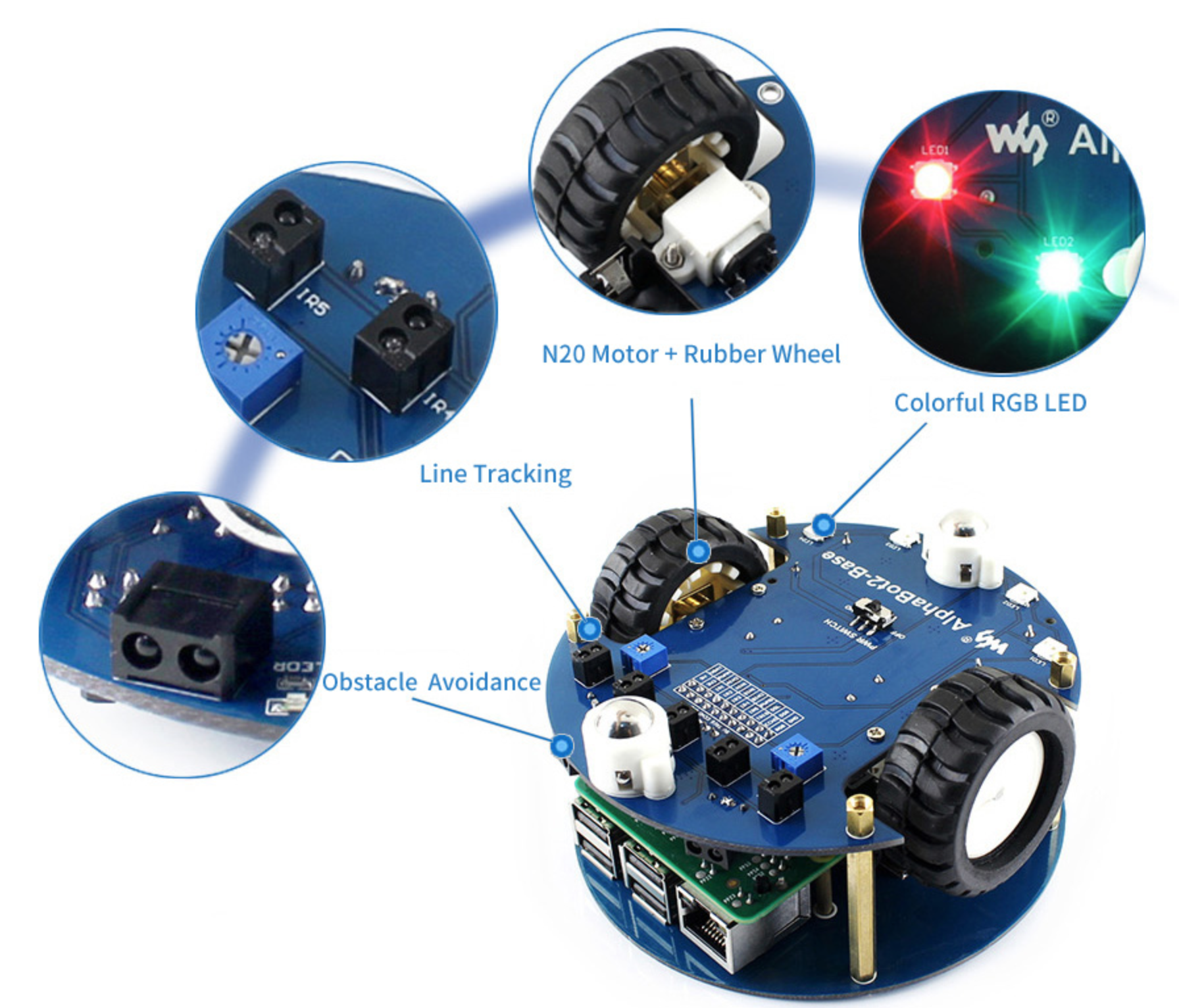}
  \end{center}
  \caption{\footnotesize Demonstration kit - simple wheeled robot with Raspberry PI and a camera sensor. The robot is able to navigate, avoid obstacles, and track lines.}
  \label{fig:lego}
  \vspace{2mm}
\end{figure}

\noindent\textbf{\textit{Demonstration Kit \& Emulators:}} 
The suggested demonstration kit includes {a simple low-cost wheeled robot with Raspberry PI and a camera sensor} (See Figure \ref{fig:lego}) \cite{demorobot}. The main component of our robot kit is a {Raspberry PI}, which runs several {Linux distributions} and can integrate with common AI and robotics software frameworks such as {Robot Operating System (ROS)}, {Tensorflow}, {PyTorch}, and {OpenCV}. This programmable robot will allow us to include various demos and student labs aimed at improving their understanding of concepts. 

A free alternative, can also be developed using an equivalent {robot emulator} built on the open-source {Gazebo} software \cite{koenig2004}. The robot emulator can be designed with virtual camera sensing and virtual navigation capabilities in order to provide the same functionality as the physical robot in a virtual environment. The ROS code used in the course modules can be deployed in the same way for both the physical robot and the robot emulator.

\begin{figure}{h}
  \begin{center}
    \includegraphics[width=0.45\textwidth]{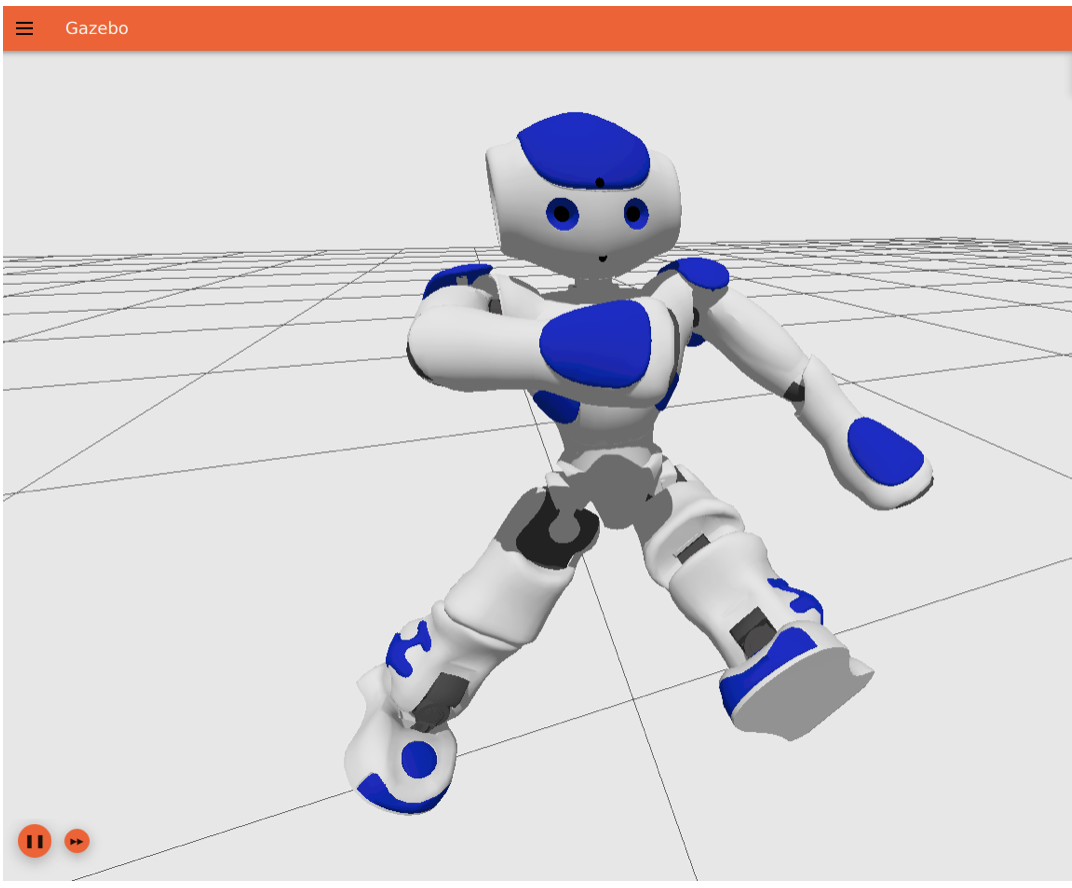}
  \end{center}
  \caption{\footnotesize Free alternate emulator \textit{Gazebo} will be used for cost prohibitive dissemination \cite{koenig2004}.}
  \label{fig:lab_space}
\end{figure}

Next, we describe each of the 7 modules in detail. 

\subsection{Module U1 - Introduction, Examples of Attacks, \& Motivation}\label{u1}


The students should first study the {anatomy of a typical robot}. Students understand that a particular robot can be divided into three abstract parts - {sensors} that accept the environmental inputs, {an AI framework} that processes these inputs and compute certain outputs that are implemented in the physical world by {actuators}. Various examples and different type of robot systems can be introduced ranging from household robots, warehouse robots, surgical robots, drones, construction robots, etc. This is where the students can be introduced to the{ simple wheeled robot with Raspberry PI, the Gazebo emulator, and a camera sensor} \cite{demorobot}. {Raspberry PI} runs several {Linux distributions} and can be integrated with {Robot Operating System (ROS)}, a commonly used robot software framework. This robot can be used by the students for different hands-on lab assignments in this course. The Raspberry PI, along with the Robot Operating System (ROS), have well documented Application Programming Interface (APIs) and a history of long-term support, making it ideal for student lab assignments, development activities, and our dissemination activities. The cost of the simple wheeled robot system (approximately \$170 per unit) is also conducive to its adoption in different universities and community colleges. 
The first lab assignment, {Lab U1.1} should involve coding and students learning programmatic interactions with the simple wheeled robot with Raspberry PI and a camera sensor.

This module will also make students {understand the risk of cyber-attacks against robotic systems}, and at the same time highlight some recent attacks \cite{eykholt2018robust,turtlerifle,robottoddler,aiimage}. Examples can include interaction of robots and equipment with larger cyber-physical ecosystems \cite{sontowski2020cyber, gupta2020security,narayanan2016obd_securealert, narayanan2016using,ramapatruni2019anomaly}. We impress upon students that parts of a robot described above, are all susceptible to cyber attacks. One can perform in front of the students - {demo attacks on all of the three parts of a robot, sensor, AI framework, and actuator}. In the second lab, {Lab U1.2}, students can simulate a simple cyber attack on the simple wheeled robot with Raspberry PI and a camera sensor system. This will involve running an attack on the Gazebo emulator as well. The aim here is to allow them to see in person some of these attacks in action so as to improve their understanding about these concepts. Details about the attacks and security strategies to protect the robotic subparts will be covered in Modules U3, U4, U5, and U6. 
Another important idea that can be included in this module is the understanding about {the social and economic implications of compromised robotics systems}. Students can be informed using real world examples that cyber attacks against robots, that they have crafted, can have very real consequences and it is their responsibility as robot developers to ensure their systems are protected against various cyberattacks.

\subsection{Module U2 - Robotic AI attack surfaces and penetration testing}\label{u2}


After introducing the students to robotics concepts and various attack demos in U1 (Section \ref{u1}), we use this module to raise awareness about common {robot focused attacks}, {attack surfaces} and {penetration testing techniques}. 
We begin this module by highlighting {broad attack surfaces} on a robot such as, {hardware components} (hardware trojans \cite{5406669}, backdoors, unauthorized access \cite{6378199}), {firmware related attacks} (worms \cite{7929597,falliere2011w32}, ransomware \cite{mohurle2017brief}, botnet \cite{feily2009survey}, reverse engineering \cite{goyal2012obfuscation}, code injection \cite{7929597}, etc.), {robotic communication} (jamming, de-authentication, traffic analysis, eavesdropping, false data injection, denial of service, replay, man in the middle, etc.) \cite{ahmad2018analyzing}, and finally moving {the focus to AI specific attacks}. Our discussion on {AI specific attack surfaces} will first introduce students to the {MITRE ATLAS (Adversarial Threat Landscape for Artificial-Intelligence Systems) knowledge base} and other efforts \cite{mitreatlas,piplai2020knowledge} modeled after the MITRE ATT\&CK framework \cite{mitreattack}. ATLAS includes a well-defined overview of adversary tactics, techniques, and case studies for AI systems based on real-world observations, demonstrations from AI security groups, and from academic research. Discussion can incorporate elements of robotics related Cyber Threat Intelligence \cite{neil2018mining,sills2020cybersecurity, mittal2016cybertwitter,khurana2019preventing,ranade2021generating,mitra2021combating}. Describing ATLAS will stimulate a discussion about AI specific attacks including {input sensor attacks, robotic AI training time attacks, inference attacks on robotic AI models, and actuator-based attacks}. We will expand each of these with Tactics, Techniques and Procedures (TTPs) followed by demos and examples. This will help students understand the basics of U3 (Section \ref{u3}), U4 (Section \ref{u4}), U5 (Section \ref{u5}), and U6 (Section \ref{u6}) that {delve deeper}. 

Another concept that will be covered is {basic penetration testing techniques to simulate AI attacks and check for exploitable AI vulnerabilities}. To provide hands on experience with penetration testing we can use the recently developed {open-source toolkit} {Counterfit} \cite{counterfit}. Counterfit, developed by Microsoft allows penetration testing and red teaming of AI systems. The toolkit contains published attack algorithms that can be used for red team operations to evade and steal AI models. The students will use Counterfit against a compromised robotic AI model in {Lab U2.1} to get a better understanding of penetration testing against robotic AI systems. {Counterfit} {directly supports} the {MITRE ATLAS framework} enabling students to better understand both. {Lab U2.2} will involve students mapping their Lab U2.1 outputs to MITRE ATLAS. 

\subsection{Module U3 - Attack patterns and security strategies for input sensors}\label{u3}


This module focuses on introducing the students to {input sensors and relevant attack patterns}. We begin by introducing different types of hardware sensors like, {visual cameras, hyperspectral cameras, sonar, radar, and lidar}. Our discussions focuses on allowing the students to understand the {inner workings} of these sensors and how each is {vulnerable to adversarial attacks} \cite{goodfellow2015}. For example, visual cameras work by measuring light intensity and can fail when a light source is compromised or when dealing with adverse lighting conditions \cite{porav2018}. Students will be introduced to software vulnerabilities in programs that directly process data coming from sensors such as {OpenCV} \cite{opencv2000}, {Tensorflow Image} \cite{tensorflow2015}, and {Open3D} \cite{open3d2018}. These frameworks have certain limitations and vulnerabilities; and often make assumptions about the structure and format of sensor data and can fail when these assumptions are not met \cite{goodfellow2015}. 
In the first {lab U3.1}, students are given the opportunity to experiment with a {physical camera} mounted on a mobile robot and paired with the {open-source computer vision library, OpenCV} \cite{opencv2000} to process the sensor data. The experiments will be conducted by running a pre-programmed obstacle detection algorithm in OpenCV \cite{nubert2018}. Students {observe how the obstacle detection performance degrades under different lighting conditions}. Students also {design an adversarial input} (e.g. an obstacle with a transparent surface or a fake picture of an obstacle) that can fool the detection system.

Next, the students are introduced to current state of the art {security strategies} against input attacks. These include (i) {verification/sanity checking} (ii) {sensor array with built-in redundancy} and (iii) {redundancy with multi-modal sensors}. Students are given the opportunity to visualize each type of defense strategy in terms of how they introduce redundancy to the overall sensor data processing system and help mitigate attacks. In {lab U3.2}, students are tasked with writing code to {perform verification and sanity checking of input sensor data}. The OpenCV library provides several options for the students to analyze the image properties including color-space conversions, histogram of pixel intensities, and feature points computation \cite{opencv2000}. The students {develop image processing algorithms to identify situations} with ambiguous objects or adverse lighting conditions {where the input sensor data should not be trusted} by the robot.



\subsection{Module U4 - Training attacks and associated security strategies}\label{u4}


Modules U4 and U5, focus on attacks on {AI-based components} of robotic systems. 
Modern robots contain multiple software components such as object recognition, voice commands, and facial recognition, that are build using an AI pipeline and these components can be {vulnerable to} (i) {training attacks} (U4) and (ii) {inference attacks} (U5). To be able to successfully defend against these types of attacks, students first need to gain {familiarity with the inner workings of these models}, learn about how the training process works, learn about how inference is carried out, and {understand the common attack vectors} and how they can be prevented. This module aims to help students understand how {vulnerabilities are enabled by inherent limitations} underlying these algorithms, and how {data can be weaponized} in ways which require an extension of how we model cyber adversary behavior, to reflect emerging threat vectors and the rapidly evolving adversarial attack lifecycle. In {lab U4.1}, students study the {anatomy of a typical AI pipeline} in robotics and implement a simple navigational AI model using the {Tensorflow framework} \cite{tensorflow2015}. Students will build a {language model based on GPT-2 \cite{radford2019} to interpret voice commands directed at a robot}. Students get to {experience the training process} by experimenting with different text examples which serve as input to the robot's language model. Next, {students are able to carry out inference} with the trained model to get the robot to recognize different voice commands that perform navigation actions such as moving forwards, rotating in place, etc.

After this, students learn about how the {training process of these AI models can be compromised through model poisoning}. Students learn about how {the training data} can be tainted by {label corruption, tainted data from open-source datasets, or chaff data from the acquisition process}, leading to an AI model that would function unexpectedly or is unable to recognize the current input data \cite{barni2019}. Students are exposed to several well-known model poisoning cases like, {Microsoft Tay chatbot incident} \cite{tays2019} and the {VirusTotal poisoning incident} \cite{virustotal2021}. With the Microsoft Tay chatbot incident, students study how adversarial training samples can cause undesired behavior. With the VirusTotal poisoning incident, students learn how mutant samples uploaded to the training platform can fool an AI model into incorrectly classifying an input. Finally, students also study defense strategies against model poisoning attacks such as {data cleaning}, {ensemble models}, {model retraining}, and {federated learning} \cite{khurana2019preventing}\cite{zhao2020pdgan}. In {lab U4.2}, students {study the effects of model poisoning} on a robotic AI model. Students {construct bad training examples} as input to the AI model (from {lab U4.1}) and {observe how the navigational task gets impacted} with changes to the training data. Students analyze various factors such as the {source, quantity and quality of training data} and reason about how this {effects the reliability or trustworthiness of an AI model}.



\subsection{Module U5 - Inference attacks and associated security strategies}\label{u5}

After understanding training attacks on AI components of robotic systems, students next study {inference attacks on robotic systems}. These inference attacks include {model stealing}, {model evasion}, and {model inversion}. Students first study and visualize the weights and layers of a neural network and understand how these contribute to computing the robot output, such as identified objects or movement commands when carrying out inference. Students gain familiarity with how these {AI models are commonly deployed on robotic systems} and {how attackers can potentially steal or exfiltrate private information or intellectual property} {via their inference APIs}. Students learn about adversarial techniques such as {Simple Transformation} of the input (cropping, shearing, translation) \cite{gao2018}, {Common Corruption} (adding white noise in the background) \cite{gilmer2019}, or {Adversarial Examples} (carefully perturbing the input to achieve desired output) \cite{goodfellow2015} which an attacker may use as a {model evasion} tactic to prevent correct output computation. Students next learn about {model stealing attacks} where the attackers are able to build a shadow model whose fidelity matches that of the victim by exploiting the robot's inference engine. Finally, students learn about {model inversion attacks} where by querying the robot's inference engine strategically, an adversary could extract potentially private information embedded in the training data. In {Lab U5.1}, a series of {case studies} are conducted by the students to understand how model inference attacks work. Students study several well-known model inference attack cases such as the {Cylance model evasion} \cite{skylance2019} and the {GPT-2 model replication} \cite{opengpt2019}. With the Cylance model evasion incident, students study how attackers can use logging data to understand the inner workings of the model, and then reverse-engineer the model to understand which attributes can be adjusted to cause an incorrect inference. With the GPT-2 model replication incident, students study how attackers were able to make use of public documentation about GPT-2 to recover a functionally equivalent ``shadow'' model. Students are also be assigned the task of identifying and researching other examples of model inference attacks on an AI system as well as the security repercussions of those attacks. 

Students are introduced to two forms of {defense strategies against model inference attacks}. The first form of defense attempts to {robustify} an AI model by {making its weights and layers more difficult to steal or reverse-engineer} \cite{tribhuvanesh2019}. This can be done by either online learning and readjustment of weights or by introducing bounded perturbations \cite{tribhuvanesh2019}. The second form of defense takes a {hybrid approach} by using AI algorithms primarily to handle the unknown quantities, but adding an additional layer of verification with tried and tested algorithms used in legacy systems to handle known robotics problems \cite{skylance2019}. This form of defense makes use of the {defense in depth} concept. Finally, in {Lab U5.2} students are assigned {a coding exercise to implement one of these security strategies} on the AI models from U4. Students then verify and compare the performance of these secured models when faced with model inference attacks.



\subsection{Module U6 - Actuator attacks and associated security strategies}\label{u6}


This module focuses on introducing the students to {robot actuators and common attack patterns} that are able to bypass the robot's input sensors as well as inner computation model and {specifically target the actuators}. The students are {introduced to different types of hardware devices} that are used by robotic AI systems to perform actions in the surrounding environment such as {wheels} for mobile robots, {grippers} for humanoid robots, and {surgical instruments} for medical robots. Students study potential attack surfaces for robot actuators such as {firmware, communication protocols, and network components} \cite{bonaci2015}. Students also learn about how to improve security features of {robotics middleware} like the Robot Operating System (ROS), by incorporating authentication, encryption, and process profile features \cite{radford2019} to ensure that robot actuators are adequately secured. To focus on {AI specific defenses}, {AI enabled digital twin based, predictive anomaly detection models} will be discussed \cite{narayanan2016obd_securealert}. In {lab U6.1}, students get to experience actuator attacks in the form of a {misbehaving robot} by going through the process of building a mobile robot that moves along a specified path and then {injecting malicious code to the actuator} that causes the robot to misbehave and deviate from its original path. 

Next, students are introduced to the associated {security strategies} for robot actuators to sanitize the control signals that are sent to the actuators and ensure that the robot is operating safety and securely in the physical world. Students learn about {access control} methods to only allow verified programs to send control signals to the actuator driver and {actuator action verification} methods to check that the actuator actions are correct and comply with the intended robot behavior \cite{carvalho2016}. {Creation of anomaly detection models} will also be covered. Students are taught to gather `normal' robotic operation data and use time series modelling techniques to generate alerts for anomalous states that can be useful for the person operating a robot will be discussed \cite{narayanan2016obd_securealert,banerjee2017generating,sai2021,chukkapalli2021privacy}. In {lab U6.2}, students implement {actuator action verification} to address the problem of the misbehaving robot from the previous lab. Students implement rule-based actions in the actuator code to prevent the robot from carrying out unwanted actions.



\subsection{Module U7 - Ethics of AI,  robotics, and cybersecurity}\label{u7}




The intended goal of this module is to help students understand that AI, robotics, and their security will have a {significant impact} on the development of humanity. These have prompted fundamental questions about {privacy and surveillance, manipulation of behaviour, opacity of AI systems, bias, human-robot interaction, employment, and machine ethics} \cite{platostanford}. Each of the fundamental questions will be discussed with the students along with various examples and case studies. Special attention will be paid to the {concept of AI trustworthiness} \cite{nsffai} which, in turn, depends on the ability to assess and demonstrate the {fairness} (including broad accessibility and utility), {transparency, explainability, impartiality, inclusivity}, and {accountability} of such systems. Students are made to understand specific national and international {laws and regulations} that can come into play while building secure robotic systems \cite{california, gdpr}. Concrete examples linking the aforementioned fundamental questions with existing laws are discussed. {Lab U7.1} requires students to investigate and research a robotic cyber incident. Here they are tasked with explaining the cyber incident, research relevant legal statutes, and give their opinions. 

Afterwards, we impress upon the students the importance of {human-robot interaction (HRI) studies} \cite{8625758, Patrick2002PrivacyT, vitale2018more,kelly2012privacy}. These studies enable robotic developers to {understand user needs} with respect to {data collection and processing, information manipulation, trust, blame, informational privacy, and security}. We explain to the students how to set up a scientifically sound privacy and security HRI study, collect data, gather inferences, and use the results to make informed decisions while developing robots and its AI subsystems. {Lab U7.2} will entail students executing a comprehensive privacy and security HRI study on a robotic system of their choice. Students create a security/privacy related HRI experiment, an Institutional Review Board (IRB) proposal, along with the necessary pre- and post-surveys. 


\section{Conclusion}\label{conc}
In this article, we provide an outline of a course with an aim of inculcating a culture of preparedness against AI security threats to pervasive robotic systems. We firmly believe that such a course when introduced at various universities and educational institutions will produce graduates and future workforce which will be well versed and equipped to prevent, detect and mitigate against sophisticated cyberattacks.

\section*{Acknowledgement}

This work was supported by a grant from the National Science Foundation. 

\bibliography{references}

\end{document}